\documentclass[12pt,preprint]{emulateapj}

\shorttitle{Hunting the Coolest Dwarfs}
\shortauthors{Schneider et al.}

\begin{document}

\title{Hunting the Coolest Dwarfs: Methods and Early Results}

\author{A. Schneider}
\affil{Department of Physics and Astronomy, University of Georgia,
    Athens, GA 30602}
\email{aschneid@physast.uga.edu}

\author{Carl Melis}
\affil{Center for Astrophysics and Space Sciences, University of California, San Diego, CA 92093-0424}
\email{cmelis@ucsd.edu}

\author{Inseok Song}
\affil{Department of Physics and Astronomy, University of Georgia,
    Athens, GA 30602}
\email{song@physast.uga.edu}

\and

\author{B. Zuckerman}
\affil{Department of Physics and Astronomy, UCLA, Los Angeles, CA 90095}
\email{ben@astro.ucla.edu}

\begin{abstract}
We present the methods and first results of a survey of nearby high proper 
motion main sequence stars to probe for cool companions with the Gemini 
camera at Lick Observatory.  This survey uses a sample of old (age $>$ 2 Gyr) 
stars as targets to probe for companions down to temperatures of 500 K. 
Multi-epoch observations allow us to discriminate comoving companions from 
background objects.  So far,our survey successfully re-discovers the wide T8.5 companion 
to GJ 1263 and discovers a companion to the nearby M0V star \objectname{GJ 660.1}.  
The companion to \objectname{GJ 660.1} (GJ 660.1B) is $\sim$4 magnitudes 
fainter than its host star in the J-band and is located at a projected separation 
of $\sim$120AU.  Known trigonometric parallax and 2MASS magnitudes for 
the GJ 660.1 system indicate a spectral type for the companion of M9 $\pm$ 2.  
\end{abstract}


\section{Introduction}
The unique spectroscopic properties of the lowest luminosity brown dwarfs led to the introduction 
of two new spectral classes, L dwarfs and T dwarfs (\citealt{kir05}).  Some wide-field 
searches for low temperature T dwarfs and Y dwarfs are currently in progress, 
and several objects with T$\sim$500-700 K have been discovered (\citealt{war07}, 
\citealt{del08}, \citealt{burn09}, \citealt{eis10}, \citealt{luc10}, \citealt{del10}, 
\citealt{leg10}, \citealt{mai11}, \citealt{bur11}). Although not fully characterized, 
the recent discovery of two cool brown dwarfs may indeed have been the first 
examples of the Y-type (\citealt{liu11}, \citealt{luh11}, \citealt{rod11b}).  Additionally, seven ultracool
brown dwarfs have recently been discovered with the Wide-field Infrared Survey
Explorer (WISE), six of which are confirmed spectroscopically to have spectral types
of Y0 or later (\citealt{cus11}).  Characterizing these Y-dwarfs will be challenging
because it is very difficult to obtain reliable physical parameters such as age, distance, 
and metallicity, for such objects.  Lowest luminosity brown dwarfs discovered as companions
of stars, however, will have better constrained physical parameters since they can be 
inferred from the host star.  For this reason, any found companion can be an 
important benchmark for use in the study of substellar objects.  To this end 
we have initiated an imaging survey to identify the coolest substellar objects as 
companions to nearby stars.  

\section{Role of Age in Search for Y-type Companions}
There will be a delay between formation of a substellar mass companion and when 
it has cooled to 500 K. The lower the mass, the faster 500 K can be reached. Thus, it 
is important to consider the masses of the lowest mass objects formed in binary systems
and accessible to imaging searches. \cite{zuc09} show that the current distribution 
of separations and masses of imaged companions with the least mass can be 
accounted for by using the minimum Jeans mass fragmentation of an interstellar 
cloud description from \cite{low76}.  They show that for a fragment to split, it must 
have at least twice the minimum fragment mass in a typical dark molecular cloud 
(derived by Low \& Lynden-Bell to be about 7 times the mass of Jupiter).  
\cite{zuc09} conclude that characteristic separations of imaged companions with 
the lowest masses (typically hundreds of AU with masses of 10-20 M$_{jup}$) 
are consistent with the Jeans fragmentation model.  Consequently, both observations 
and theory suggest that wide separation secondaries will almost always have 
masses $\gtrsim$ 10 M$_{jup}$.  We assume that companions to stars will have
masses $\geq$ 15 M$_{jup}$.

According to Baraffe et al. (2003; 2006 private communication) and \cite{bur03}, 
the time needed for a 15 Jupiter mass object to cool to 500 K is $\sim$2 Gyr. 
The relatively small number of imaged companions with masses below $\sim$15 M$_{jup}$
implies that the detection of a Y-type companion with an age $<$ 2 Gyr will be very
rare.  Therefore, relatively older systems will make better targets to search for cool companions.  
Brown dwarf secondaries having masses larger than 15 M$_{jup}$ are more common than the 
lowest mass companions listed in Table 1 of \cite{zuc09} (see Table 2 of same 
paper). These companions would have to be around stars older than 2 Gyr in 
order to cool to the 500 K effective temperature that might herald the onset of the 
Y-class.  According to the models of \cite{bar03}, even brown dwarfs as old as 7 Gyr 
would have to be less massive than $\sim$25 M$_{jup}$ to have cooled to 500 K.  
Therefore, Y-dwarfs themselves must be relatively low-mass.

\section{Sample Selection}
We are interested in probing for Y-type dwarfs which are born with separations 
from their host star in the hundreds to thousands of AU range. It is for the above 
reasons that the targets selected for our proper motion companion search are 
those nearby stars with ages $>$ 2 Gyr.  Our current list of 50 targets consists 
of stars from the Gliese catalog because these stars tend to be old ($\gg$ 100 
Myr, \citealt{song03}).  Stars with the highest proper motions (total proper motion 
$\geq$ 650 mas/yr) were selected from this catalog because of instrumental 
considerations (see Section 4).  Each potential target was cross-matched with the 
ROSAT All-Sky Survey (RASS) with a search radius of 2' to check for X-ray emission 
as a possible indicator of age.  Of the targets in our sample, seven were found to have  
been detected in X-rays.  The X-ray luminosity measured for each still indicates an 
age at least as old as that of the Hyades (age $\sim$600 Myr).  X-ray luminosity upper 
limits were calculated for the remainder of our targets.  Detections and upper limits 
for our targets are shown in Figure 1.  Additionally, space motions were calculated 
to verify that they do not fall within the young star UVW defined by \cite{zuc04}. These 
regions are areas of UVW space inhabited by young (age $\lesssim$ 100 Myr), nearby stars.  
UVW space motions for our targets are plotted in Figure 2, which shows that the 
space motions of our targets are inconsistent with the space motions of nearby 
young stars.  Each target was also checked for any measure of binarity to exclude 
those with known companions.  

Our target list and current observation status are given in Table 1.  Parallax 
measurements are from the Gliese catalog. Right ascension, declination, 
and J-magnitudes are from the Two Micron All Sky Survey (2MASS). 
Spectral type and proper motion are from the SIMBAD astronomical database.  
Distributions of proper motion and spectral type are shown in Figure 3. After the 
target list was created, we then checked each target for GALEX near ultraviolet 
emission as a possible additional indicator of age (\citealt{shk11}, \citealt{rod11}).  
Twenty-eight of our targets had corresponding GALEX matches.  Comparison with
 young stars (\citealt{zuc04}), Hyades members (obtained from the WEBDA open
 cluster database), and the rest of the Gleise catalog again indicate 
 that our sample consists of an older population (Figure 4).   

\section{Observations}
Observations for this survey began in August of 2007 and we are still actively 
acquiring second epoch data for selected stars.  This survey is being conducted 
with the Gemini camera (McLean et al. 1993) located on the Shane 3m Telescope 
at the Lick Observatory, Mt. Hamilton, CA. The Lick Gemini camera is unique for 
our purposes with its combination of a wide field of view ($3'\times3'$) and a coronagraphic 
spot.  Our observations of main sequence F-, G-, K-, and M-type stars employ this 5'' 
coronagraphic spot to best suppress scattered light and obtain maximum sensitivity in 
the full field of view. These observations yield a radial field of view of $\sim$90'' and 
allow us to probe separations out to 900 AU at 10 pc and 1800 AU at 20 pc (distances to 
our targets range from 10 - 25 pc).  While the smallest separations we can probe depend
on the apparent brightness of the target star, the typical smallest separations to detect 
a source at 3$\sigma$ above the background are $\sim$10''.

Based on T dwarf spectra and theoretical models, the very low temperature objects we 
are seeking are anticipated to be more easily detected at J-band rather than K-band. 
Therefore, all targets are being observed in Gemini J-band imaging.  With 30 minutes 
of on-source integration time per object, we can (in good conditions) achieve a limiting J-band magnitude of 20 
at the 6$\sigma$ detection level. Based on the Baraffe et al. (2003) models, this magnitude 
limit is sufficient to detect companions as old as 5 Gyr with temperatures down to 500 K 
out to 15 pc, and companions with temperatures down to 600 K out to 25 pc.   

We search at two epochs for co-moving companions to old, high proper motion, 
main-sequence stars.  Through the end of 2009, we have obtained first epoch images 
for 50 main sequence stars with spectral types from late-F to mid-M; second epoch 
imaging to similar depths have been obtained for 41 of these stars.  All target images 
were reduced with custom in-house IDL software routines. Science frames were reduced 
by performing sky subtraction, flat division, bad pixel masking, shifting, and averaging.  
Source extraction for this project was done using the IRAF routine DAOFIND. 

Background 2MASS sources in each field can be used to apply a world coordinate 
system (wcs) to each image.  This is accomplished by matching pixel coordinates with 
2MASS coordinates and using the IRAF task CCMAP.  Once the transformation is found, 
a wcs can be applied to the image with the iraf task CCSETWCS.  Once a wcs is applied 
to an image, measuring separations and position angles can be performed with the IDL 
routines $gcirc.pro$ and $posang.pro$, both part of the IDL Astronomy User's Library.   
This process also allows us to calculate plate scale information for each image.  We measure 
a GEMINI pixel scale of 0.70'' per pixel for these observations.  

\subsection{Astrometry}
The proper motion of a particular target can be measured and compared to any motions 
exhibited by stars in the surrounding field by imaging the area around the target star at 
two epochs separated by a sufficient interval of time. The IRAF task GEOMAP can create 
a general transformation between two sets of coordinates corresponding to sources in 
the same field at two different epochs. GEOMAP uses a polynomial fit to the sets of 
coordinates to account for translation, rotation, scaling, and distortion (we use the default 
setting of the polynomial fit, which is a power series in x and y of order 2).  Residuals of 
the pixel shifts from one epoch to the next can be used to flag any high proper motion 
objects against the near zero proper motion of background stars.  The scatter in the 
residuals of the background stars can also be used to provide a measure of our astrometric 
measurement uncertainty in the following manner.

Firstly, we measure the center position of background star offsets by calculating the average 
offset in the x and y directions using the pixel shifts from one epoch to the next.  Then, 
from this position, we calculate the standard deviation of residuals for all background 
stars.  This standard deviation is typically between 0.1 and 0.2 pixels for our observations.  
Since our targets are high proper motion stars ($\mu$ $>$ 650 mas/yr), in one year, a 
typical target displays a positional displacement of $\sim$1 pixel with respect to background 
stars.  Therefore, co-moving companions to our targets should be detectable at the 
5-10$\sigma$ level by two epochs separated by a year. 

\section{Early Results} 
For each observation, we calculate a 90\% magnitude completeness limit.  Using 
these limits, known parallax measurements, and the "COND" evolutionary models 
from \cite{bar03}, we estimate the minimum mass and effective temperature of a 
companion that could be detected in each image.  The mass and corresponding 
effective temperature sensitivity limits for each target that has two observed epochs 
of good quality are shown in Figure 5 (assuming an age of 2 Gyr).  The average 
temperature sensitivity limit for these pairs of images is $\sim$600 K.  This corresponds 
to an average mass sensitivity limit of 0.018 solar masses ($\sim$19 Jupiter masses) 
for an estimated age of $\sim$2 Gyr. One of the coolest dwarfs with a measured 
parallax is the T9$+$ dwarf UGPS 0722-05 discovered 
by \cite{luc10}.  UGPS 0722-05 has an absolute J-band magnitude of 18.5$\pm$0.2.  By
comparing this with our magnitude completeness limits, we would be able to detect an 
object as faint or fainter than UGPS 0722-05 around 20 of our targets (40\%).  

The capability of GEOMAP to identify faint, co-moving sources with only one year 
time-baseline between epochs is demonstrated with our imaging of the 
\objectname{GJ 1263} (Wolf 940; \citealt{burn09}) wide binary system (M4 + T8.5).  
We first observed \objectname{GJ 1263} on 2008 August 10 and again on 2009 
August 28.  Its co-moving companion was identified by plotting positional offsets of 
surrounding objects (Fig. 6) in the method described above.  The 1$\sigma$ uncertainty 
in position for background objects of median 2MASS J-mag = 16.5 is $\sim$ 0.10 pixels.  
With a total offset of $\sim$ 1.213 pixels (the total proper motion of GJ1263 is 
$\sim$970 mas/yr), the candidate companion is $\sim$12$\sigma$ away from the 
center of all background source residuals.  Derived properties of GJ 1263B are 
displayed in Table 2.  Absolute J magnitudes, separations, and projected separations 
are in good agreement with those quoted in \cite{burn09}.       

The first new discovery of our survey is a co-moving companion to the star 
\objectname{GJ 660.1}.  We observed \objectname{GJ 660.1} on 2008 August 12 
and again on 2009 August 30.  \objectname{GJ 660.1} is an M0 star located at a 
distance of $20.0^{+1.6}_{-1.3}$ pc (van Leeuwen 2007) with proper motion in right 
ascension of 189 mas/yr and -695 mas/yr in declination.  GJ 660.1 was detected in 
2MASS with J-band magnitude of 8.66 $\pm$ 0.03.  Radial velocity measurements 
(Reid et al. 1995) and the proper motions mentioned above indicate Galactic $UVW$ 
space motions (U = 0.5 $\pm$ 2.1 km s\textsuperscript{-1}, V = -52.1 $\pm$ 3.0 km 
s\textsuperscript{-1}, W = -60.3 $\pm$ 3.5 km s\textsuperscript{-1}) inconsistent with 
any nearby young stellar associations\footnote{$UVW$ are defined with respect 
to the Sun. U is positive toward the Galactic center, V is positive in the direction of 
Galactic rotation, and W is positive toward the north Galactic pole.} . The large 
negative V and large absolute value of W, along with the low position of GJ 660.1A 
on an HR  diagram and a null Rosat All Sky Survey X-ray detection imply that the star 
is older than $\sim$2 Gyr.  

Measured residuals for the companion and background sources are shown in Fig. 6.  
The 1$\sigma$ uncertainty in position for background objects is $\sim$ 0.091 pixels.  
With a total offset of $\sim$ 1.274 pixels, the companion to GJ 660.1 is $\sim$14$\sigma$ 
away from the centroid of all background source offsets.  Properties derived for this 
companion are shown in Table 2.  This companion was detected by the 2-Micron All 
Sky Survey (2MASS) along with several background sources in our images.  Residuals 
from this earlier 2MASS image (epoch 1999 March 31) confirm this companion at 
the $\sim$35$\sigma$ level.  Based on absolute measured and 2MASS magnitudes 
found for the companion and a comparison to the magnitudes of known dwarfs (using 
DwarfArchives.org), we estimate a spectral type of M9 $\pm$ 2.  Using the 
models of \cite{bar03}, the mass of this companion is determined to be between 
0.075 and 0.080 $M_\odot$ for an estimated age between 1 and 5 Gyr. This companion 
will be discussed in a future publication in more detail.

\section{Conclusion}
We report the methods and early results of a Lick/GEMINI survey of 50 old nearby 
stars for cool companions.  We illustrate the capabilities of our observing strategy 
via detection of the previously known M4 + T8.5 binary system GJ 1263AB.  In addition 
to this rediscovery, we discovered a late M dwarf companion to GJ 660.1.  By comparing 
position residuals versus those of stationary background objects, we conclude that 
GJ 660.1B is co-moving with GJ 660.1A.  We also show detectable lower limits of effective 
temperature and mass for possible companions of our targets.  Based on our detection
limits, we will be able to detect companions as faint or fainter than the T9$+$ dwarf UGPS
0722-05 for 40$\%$ of our targets.  These early results show the effectiveness of this method.

\acknowledgments

This research was supported in part by a NASA grant to UGA and UCLA.  This research 
has benefitted from the M, L, and T dwarf compendium house at DwarfArchives.org and 
maintained by Chris Gelino, Davy Kirkpatrick, and Adam Burgasser.  This research has made 
use of the WEBDA database, operated at the Institute for Astronomy of the University of Vienna.  
This research has made use of the SIMBAD database and VisieR catalog access tool, operated 
at CDS, Strasbourg, France.  This publication makes use of data products from the Two Micron 
All Sky Survey, which is a joint project of the University of Massachusetts and the Infrared 
Processing and Analysis Center/California Institute of Technology, funded by the National 
Aeronautics and Space Administration and the National Science Foundation. C.M. acknowledges
support from the National Science Foundation under award No. AST-1003318.
We thank Adam J. Burgasser for useful discussion.

\clearpage
\begin{deluxetable}{lccllllcc}
\tabletypesize{\scriptsize}
\tablecaption{Target List}
\tablewidth{0pt}
\tablehead{
 & \colhead{RA} & \colhead{DE} &  \colhead{Spectral} &  \colhead{J} & \colhead{$\mu$} & \colhead{$\pi$}  &   & \\
\colhead{GJ $\#$} & \colhead{(J2000.0)} & \colhead{(J2000.0)} & \colhead{Type} & \colhead{(mag)} & \colhead{(mas yr\textsuperscript{-1})} & \colhead{(mas)} &\colhead{Obs. Date 1} & \colhead{Obs. Date 2}
}
\startdata
 1014  & 00:35:55.57  & +10:28:35.21 & M5      &  10.22 & 1173.08 & 64.0 & 10/10/08 & ...     \\
 28    & 00:40:49.29  & +40:11:13.33 & K2Ve   &  5.69 &  758.22 & 58.5 & 8/11/08  & 8/30/09 \\
 38    & 00:51:29.64  & +58:18:07.14 & dM2     &   7.83 & 1621.69 & 53.5 & 8/12/08  & 8/29/09 \\
 1025  & 01:00:56.44  & $-$04:26:56.15 & M3.5    &  9.04 & 1325.08 & 60.0 & 10/25/07 & ...     \\
 1029  & 01:05:37.32  & +28:29:33.98 & M5      &   9.49 &  1917.83 & 79.6 & 10/9/08  & ...     \\
 52    & 01:07:07.95  & +63:56:28.43 & K7V    &   6.47 & 1578.75 & 67.2 & 8/5/07   & 8/9/08  \\
 1035  & 01:19:52.28  & +84:09:32.80 & M5      &  9.85 & 1089.15 & 73.2 & 10/10/08 & 8/31/09 \\
 87    & 02:12:20.91  & +03:34:31.09 & dM2.5   &   6.83 & 2555.85 & 86.7 & 10/24/07 & 10/8/08 \\
 3181  & 02:46:34.86  & +16:25:11.60 & M6      &  10.97 & 1012.35 & 66.7 & 10/10/08 & ...     \\
 123   & 03:06:26.76  & +01:57:53.84 & M0V    &  6.49 & 1005.56 & 63.6 & 10/24/07 & 10/8/08 \\
 124   & 03:09:03.87  & +49:36:47.94 & G0V    &   3.14 & 1265.91 & 92.4 & 10/26/07 & 10/9/08 \\
 197   & 05:19:08.48  & +40:05:56.59 & G2IV-V &  3.39 &  843.63 & 69.5 & 10/25/07 & 10/8/08 \\
 215   & 05:45:48.22  & +62:14:13.28 & K7      &  6.35 &  841.37 & 71.6 & 10/26/07 & ...     \\
 217   & 05:46:01.92  & +37:17:04.38 & K1V    &  5.83 &   705.57 & 56.7 & 10/10/08 & ...     \\
 262   & 07:03:30.44  & +29:20:15.01 & G4V    &   4.89 &    842.46 & 55.0 & 10/24/07 & 10/8/08 \\
 302   & 08:18:23.89  & $-$12:37:54.15 & G7.5V  &  4.95 & 1028.43 & 79.3 & 10/24/07 & ...     \\
 365   & 09:43:25.63  & +42:41:31.00 & K5V    &  6.09 &  828.46 & 50.5 & 10/26/07 & 10/9/08 \\
 602   & 15:52:40.53  & +42:27:05.21 & F9V    &   2.94 &  768.85 & 57.5 & 7/11/08  & 7/6/09  \\
 603   & 15:56:27.20  & +15:39:41.31 & F6V    &  3.14 & 1319.39 & 84.1 & 8/5/07   & 8/9/08  \\
 9537  & 16:01:02.65  & +33:18:12.48 & G0V    &   4.09 &  758.98 & 60.0 & 7/8/08   & 7/3/09  \\
 609   & 16:02:50.98  & +20:35:21.83 & M3      &  8.13 &  1551.35 & 99.4 & 8/10/08  & 8/29/09 \\
 651   & 17:02:36.40  & +47:04:54.03 & G8V    &  5.41 &  864.43 & 61.8 & 7/9/08   & 7/4/09  \\
 1209  & 17:04:22.34  & +16:55:55.22 & M2.5    &  8.57 &  1135.29 & 57.9 & 7/11/08  & 7/6/09  \\
 660.1 & 17:12:51.27  & $-$05:07:31.17 & M0      &  8.66 &  720.31 & 49.6 & 8/12/08  & 8/30/09 \\
 672   & 17:20:39.55  & +32:28:05.69 & G2V    &   4.16 &  1048.72 & 73.7 & 8/6/07   & 8/9/08  \\
 700.2 & 18:02:30.85  & +26:18:47.10 & K0V    &   5.55 &  717.41 & 53.0 & 8/11/08  & 8/30/09 \\
 1225  & 18:17:15.09  & +68:33:19.92 & M4.5    &  10.78 & 1730.67 & 54.3 & 7/9/08   & 7/4/09  \\
 712   & 18:22:06.71  & +06:20:37.70 & M3      &   8.67 &  1139.96 & 69.1 & 8/12/08  & 8/31/09 \\
 740   & 18:58:00.14  & +05:54:29.70 & M2V    &  6.23 & 1234.96 & 94.6 & 7/10/08  & 7/5/09  \\
 1235  & 19:21:38.68  & +20:52:02.82 & M4.5    &  8.79 & 1731.88 & 98.4 & 7/8/08   & 7/3/09  \\
 759   & 19:24:58.23  & +11:56:40.15 & G8IV   &  3.55 &  965.93 & 64.3 & 7/9/08   & 7/4/09  \\
 1248  & 20:03:50.99  & +05:59:44.02 & M1.5    &   8.63 &  935.45 & 78.9 & 7/11/08  & 7/6/09  \\
 778   & 20:03:52.10  & +23:20:26.27 & K1V    &  5.67 & 1356.05 & 55.0 & 7/8/08   & 7/3/09  \\
 1254  & 20:33:40.31  & +61:45:13.57 & M4      &   8.29 & 1061.60 & 62.5 & 8/10/08  & 8/31/09 \\
 9711  & 20:56:46.59  & $-$10:26:53.43 & dM4     &  7.76 & 1109.96 & 56.3 & 8/10/08  & 8/29/09 \\
 817   & 21:04:53.42  & $-$16:57:31.11 & M2      &  8.28 &  2233.71 & 57.5 & 10/26/07 & 10/8/08 \\
 9722  & 21:07:55.43  & +59:43:19.89 & sdM1    & 10.12 &  2109.90 & 41.7 & 8/11/08  & 8/30/09 \\
 821   & 21:09:17.41  & $-$13:18:08.03 & M3      &  7.68 & 2117.76 & 91.7 & 7/10/08  & 7/5/09  \\
 830   & 21:30:02.72  & $-$12:30:36.15 & M0V    &  6.65 & 1052.66 & 63.6 & 8/6/07   & 8/9/08  \\
 1263  & 21:46:40.40  & $-$00:10:23.35 & M3.5    &  8.36 &  969.92 & 81.8 & 8/10/08  & 8/29/09 \\
 4239  & 21:56:55.14  & $-$01:54:10.05 & dM5     &   9.88 & 1408.69 & 75.0 & 10/9/08  & ...     \\
 1266  & 22:16:20.29  & +70:56:39.53 & M2      &  8.74 &   863.38 & 44.5 & 10/26/07 & 10/8/08 \\
 1271  & 22:42:38.72  & +17:40:09.11 & M3      &  8.06 &  1220.71 & 52.6 & 8/11/08  & 8/30/09 \\
 4312  & 23:07:30.04  & +68:40:05.13 & M3.5    &  8.62 & 1145.09 & 63.5 & 10/26/07 & 10/9/08 \\
 4333  & 23:21:37.52  & +17:17:28.47 & M4      &   7.39 & 1485.52 & 92.8 & 7/9/08   & 7/5/09  \\
 4336  & 23:26:32.39  & +12:09:32.83 & M3      &   8.96 &   747.90 & 47.0 & 8/12/08  & 8/31/09 \\
 895.4 & 23:31:22.20  & +59:09:55.86 & K0V    &  5.34 & 1113.05 & 56.1 & 10/25/07 & 10/9/08 \\
 4346  & 23:35:44.45  & +41:58:03.81 & M0      &  8.10 &  717.47 & 44.0 & 8/11/08  & 8/30/09 \\
 1292  & 23:57:44.10  & +23:18:16.97 & M3.5    &    7.8 & 1465.89 & 72.1 & 8/9/08   & 8/29/09 \\
 4385  & 23:59:49.41  & +47:45:44.80 & M5      & 10.86 &  893.44 & 59.8 & 10/10/08 & ...     \\
\enddata
\end{deluxetable}

\begin{deluxetable}{lllc}
\tabletypesize{\scriptsize}
\tablecaption{Companion Properties}
\tablewidth{0pt}
\tablehead{
\colhead{Observed Quantity} & \colhead{GJ 1263 B} & \colhead{GJ 660.1 B}  & \colhead{Epoch\tablenotemark{a}}
}
\startdata
2MASS J-Mag  &   & 13.05 $\pm$ 0.05 & \\
2MASS H-Mag  &  & 12.57 $\pm$ 0.02 & \\
2MASS K-Mag    &    & 12.23 $\pm$ 0.03 & \\
Absolute 2MASS J-Mag\tablenotemark{b}   &    & 11.6 $\pm$ 0.5 & \\
Absolute 2MASS H-Mag\tablenotemark{b}   &    & 11.1 $\pm$ 0.5 &\\
Absolute 2MASS K-Mag\tablenotemark{b}   &    & 10.7 $\pm$ 0.5 & \\
Gemini J-Mag      & 18.42 $\pm$ 0.25 & 13.16 $\pm$ 0.21 & 1\\
Absolute Gemini J-Mag\tablenotemark{b}        & 17.9 $\pm$ 0.7 & 11.7 $\pm$ 0.5 & 1\\
Separation ('')     & 31.55 $\pm$ 0.01 & 6.00 $\pm$ 0.02 &  1\\
Projected Separation (AU)\tablenotemark{b}  & 395 $\pm$ 22 & 120 $\pm$ 9 &  1\\
Position Angle ($\degr$)      & 250.4 $\pm$ 0.1 & 353.1 $\pm$ 0.1 &  1\\
Gemini J-Mag      & 18.29 $\pm$ 0.27 & 12.99 $\pm$ 0.21 & 2\\
Absolute Gemini J-Mag\tablenotemark{b}        & 17.8 $\pm$ 0.7 & 11.5 $\pm$ 0.5 & 2\\
Separation ('')       & 31.64 $\pm$ 0.01 & 6.08 $\pm$ 0.02 &  2\\
Projected Separation (AU)\tablenotemark{b}   & 396 $\pm$ 22 & 121 $\pm$ 9 &  2\\
Position Angle ($\degr$)       & 250.5 $\pm$ 0.1 & 353.2 $\pm$ 0.1 &  2\\
\enddata
\tablenotetext{a}{See text for epochs.}
\tablenotetext{b}{Values and uncertainties account for distance and uncertainty in distance to each object ($12.50^{+0.75}_{-0.67}$ pc for GJ 1263B and $20.0^{+1.6}_{-1.3}$ pc for GJ 660.1B).}
\end{deluxetable}

\clearpage

\begin{figure}
\plotone{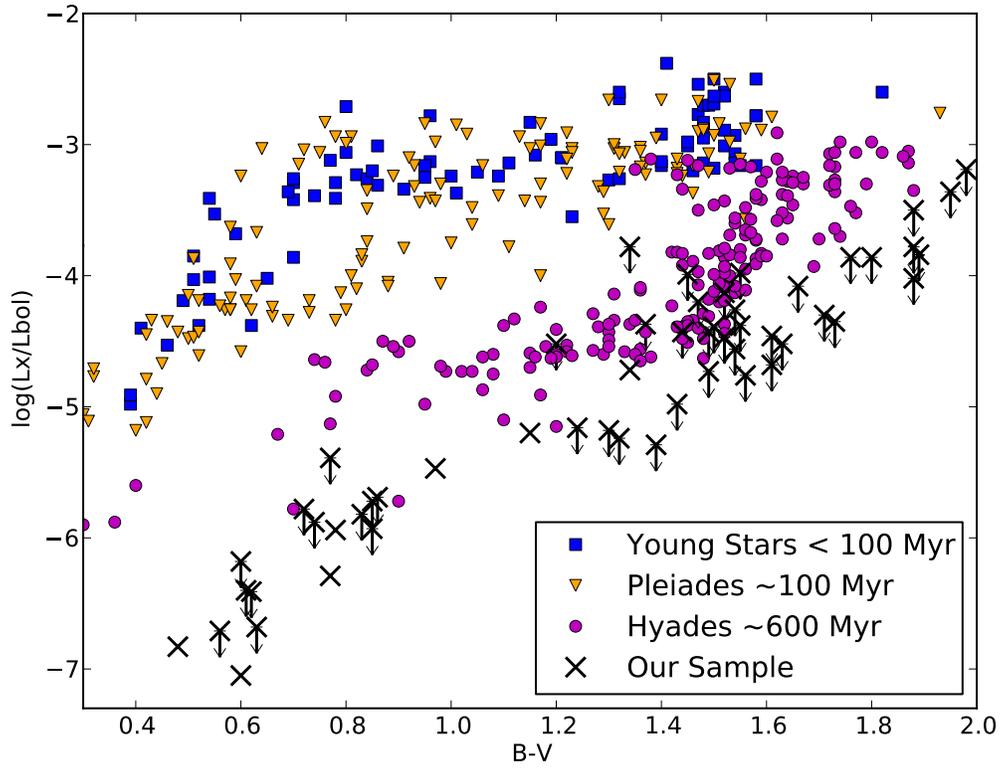}
\caption{X-ray detections and upper limits for targets in our sample with RASS detections. Crosses indicate detected sources, while downward arrows indicate upper limits.  Young stars, Pleiades, and Hyades data points are from \cite{zuc04}.}
\end{figure}
\clearpage

\begin{figure}
\epsscale{0.7}
\plotone{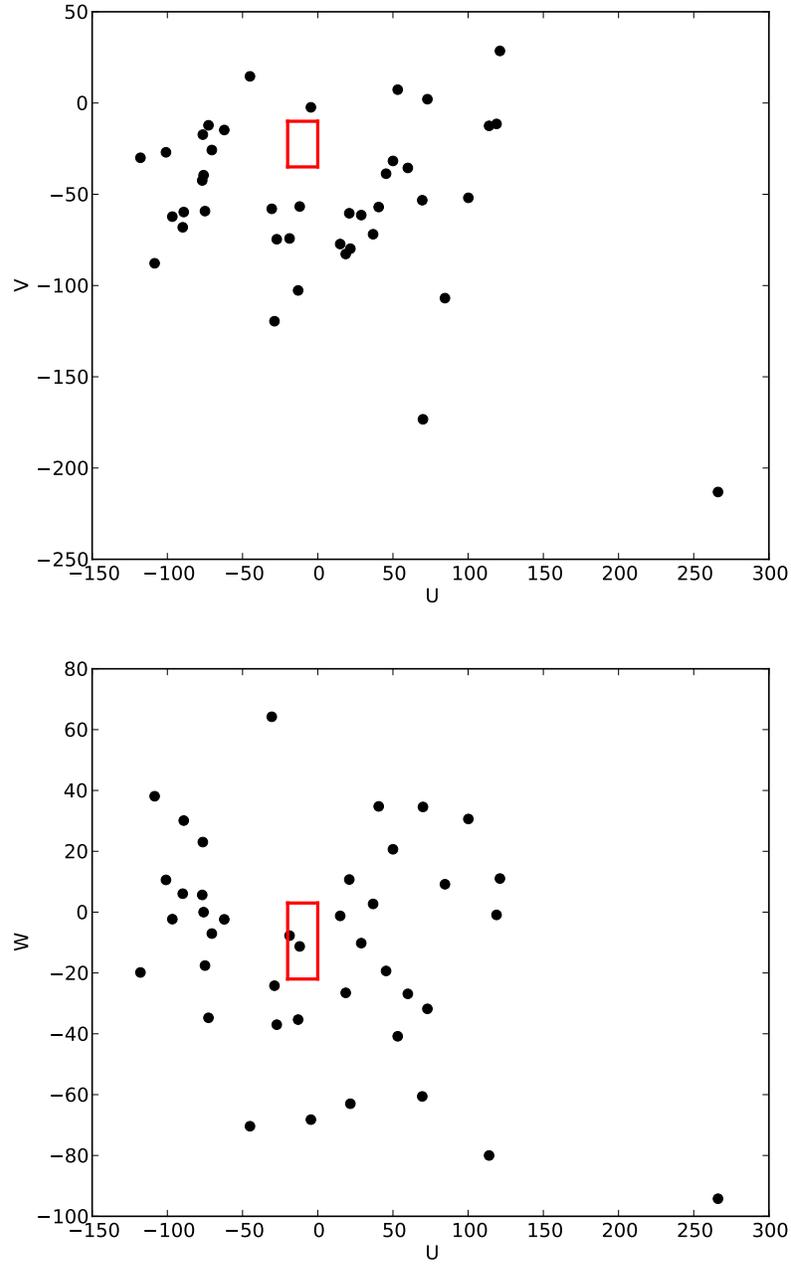}
\caption{UVW space motions for our selected targets.  The red box is the ''good box'' mentioned in the text.  For a target to be consistent with young nearby star space motion, it would need to occupy space in the ''good box'' region in both plots.  None of our selected targets do so.}
\end{figure}
\clearpage

\begin{figure}
\plotone{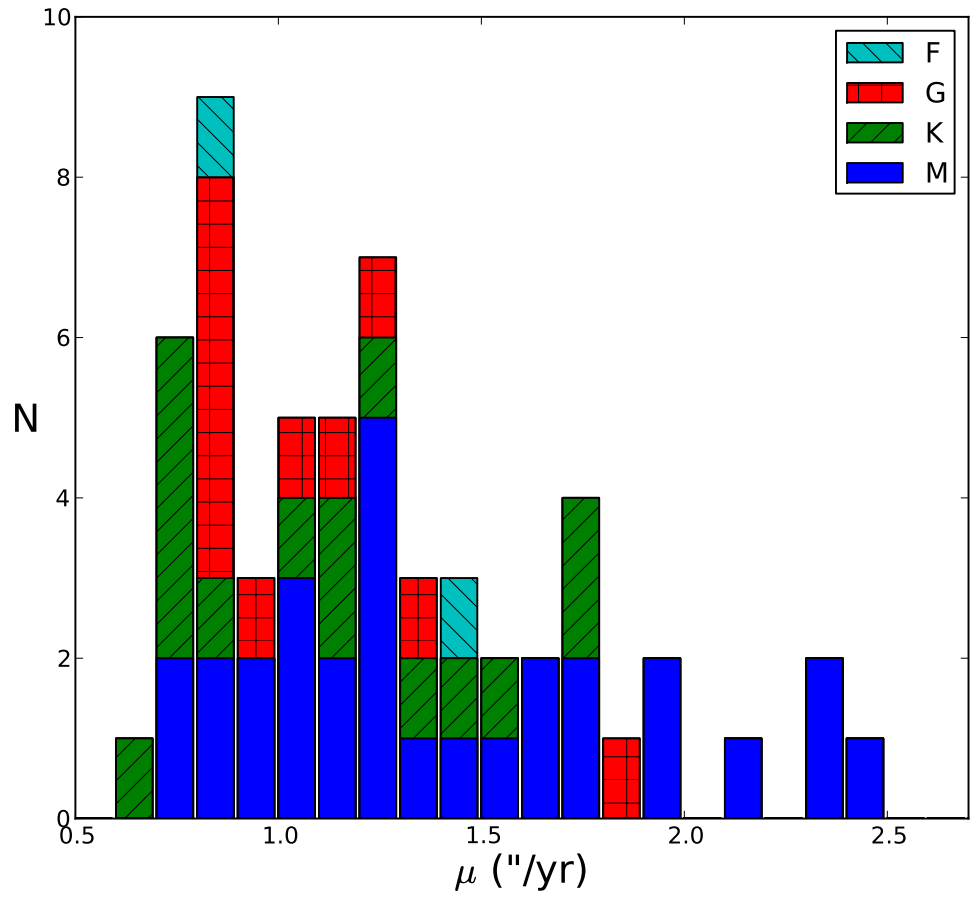}
\caption{Histogram of total proper motion ($\mu$) for the target sample.  Colors indicate spectral type.}
\end{figure}
\clearpage

\begin{figure}
\plotone{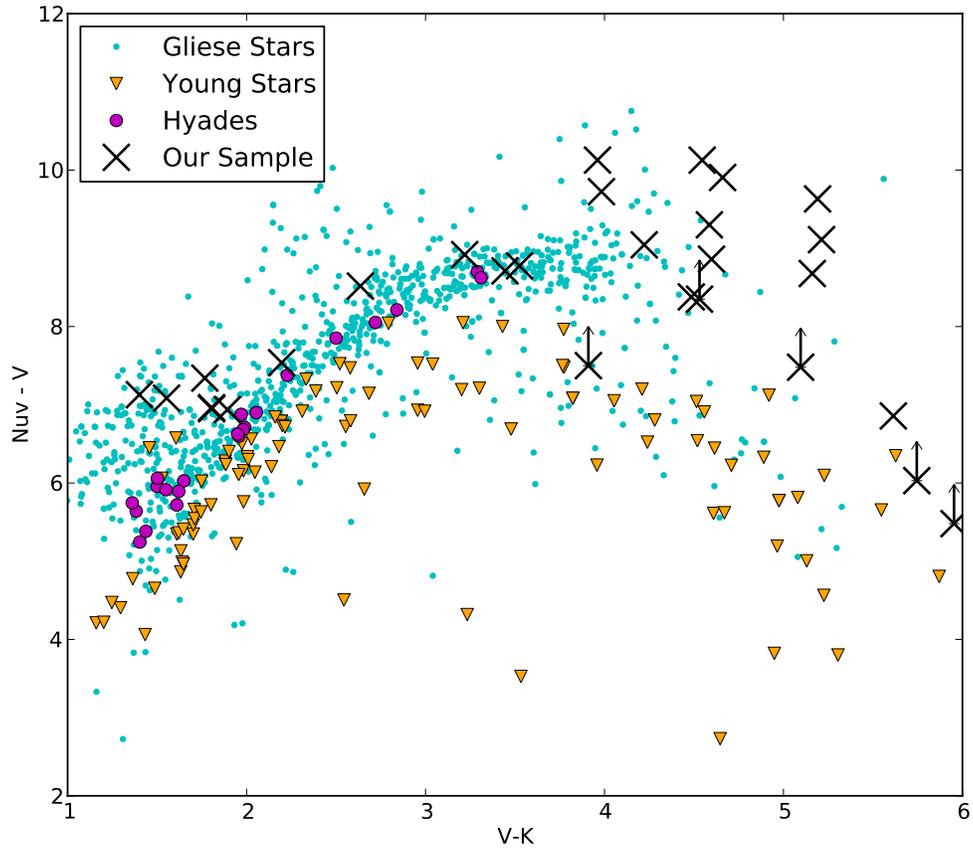}
\caption{Galex detected targets in our sample.  Upward pointing arrows indicate limits on stars without Galex counterparts within the Galex coverage area.  Young stars are from \cite{zuc04}.  Hyades members were obtained from the WEBDA open cluster database.}
\end{figure}
\clearpage

\begin{figure}
\plotone{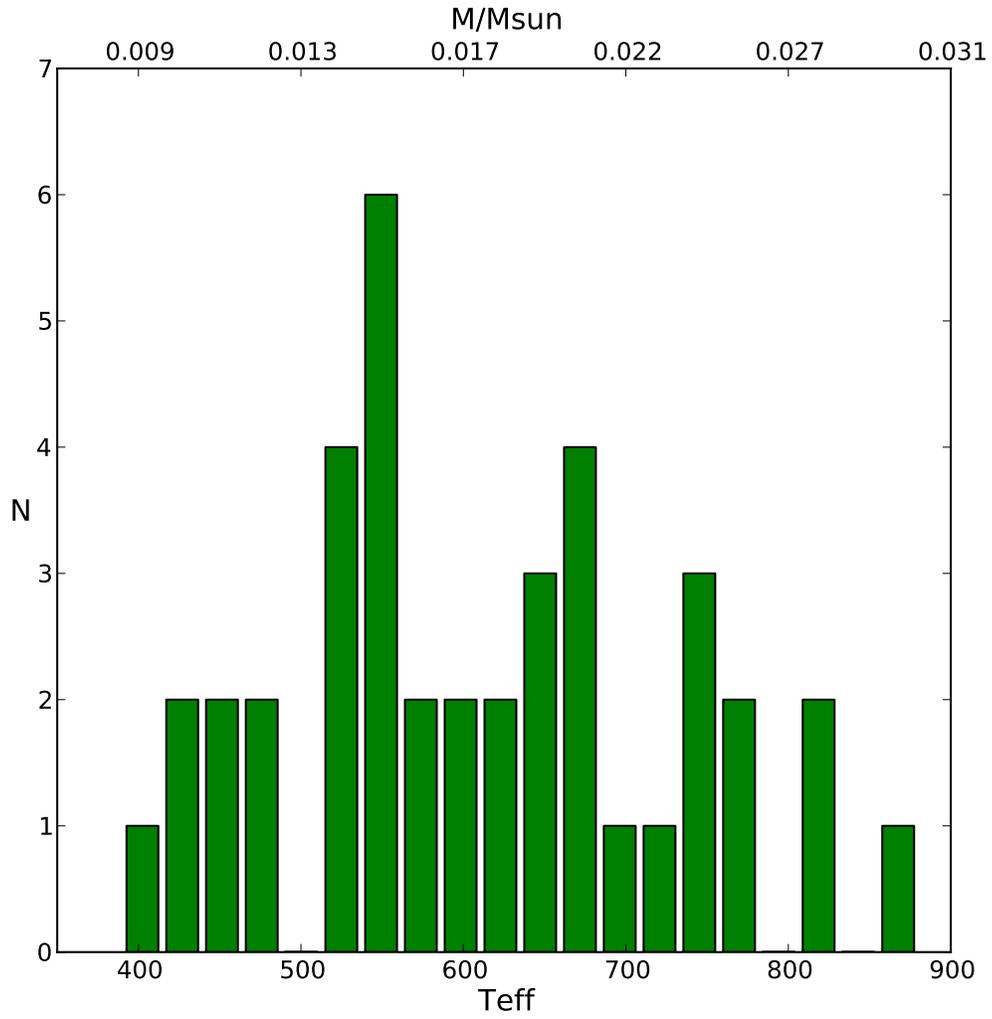}
\caption{Histogram of effective temperature lower limits for each pair of observations. Corresponding lower mass limits are shown on the upper axis.  Determination of these values is discussed in section 5.}
\end{figure}
\clearpage

\begin{figure}
\epsscale{1}
\plotone{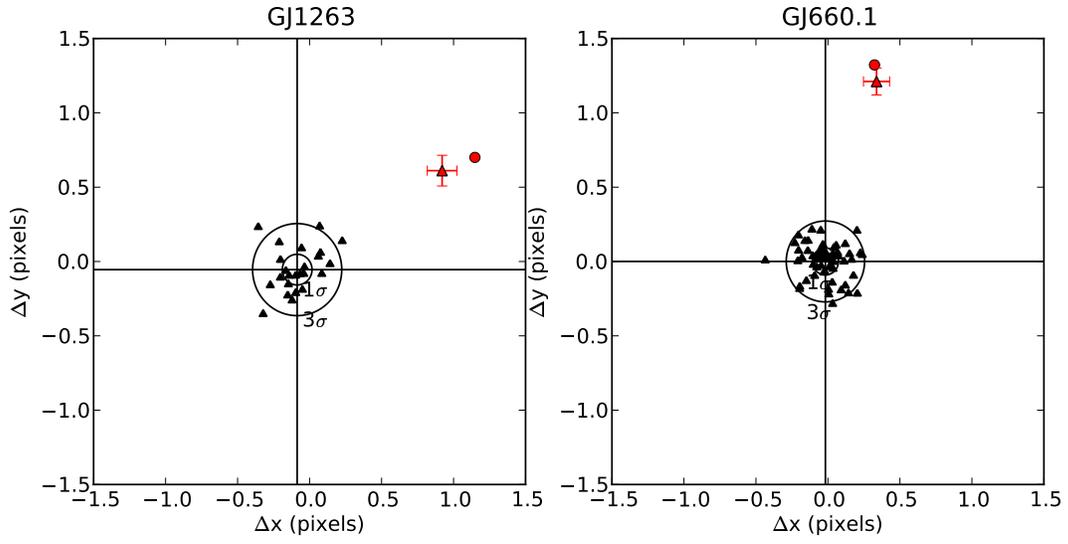}
\caption{{\it Left:} Residuals from alignment of background reference stars (black triangles) with GJ1263A (red circle) and GJ1263B (red triangle) between two epochs.  {\it Right:} Residuals from alignment of background reference stars (black triangles) with GJ660.1A (red circle) and GJ660.1B (red triangle) between two epochs.  Positions of primaries are calculated from recorded catalog positions and known parallax and proper motions.}
\end{figure}
\clearpage


\begin{thebibliography}{}
\bibitem[Ackerman \& Marley(2001)]{ack01} Ackerman, A. S. \& Marley, M. S., 2001, \apj, 556, 872
\bibitem[Baraffe et al.(2003)]{bar03} Baraffe, I., Chabrier, G., Barman, T. S., Allard, F., \& Hauschildt, P. H. 2003, \aap, 402, 701
\bibitem[Becklin \& Zuckerman(1988)]{bec88} Becklin, E.,  \& Zuckerman, B.  1988, \nat, 336, 656
\bibitem[Burgasser et al.(2008)]{bur08} Burgasser, A., Looper, D., Kirkpatrick, J., Cruz, K., \& Swift, B. 2008, \apj, 674, 451
\bibitem[Burgasser et al.(2011)]{bur11} Burgasser, A. J. et al. 2011, ArXiv:1104.2537B
\bibitem[Burningham et al.(2009)]{burn09} Burningham B. et al. 2009, \mnras, 395, 1237
\bibitem[Burrows et al.(2003)]{bur03} Burrows, A., Sudarsky, D., \& Lunine, J. I.  2003, \apj, 596, 587
\bibitem[Cushing et al.(2011)]{cus11} Cushing, M. C. et al. 2011, ArXiv:1108.4678
\bibitem[Delorme et al.(2008)]{del08} Delorme, P. et al. 2008, \aap, 482, 961
\bibitem[Delorme et al.(2010)]{del10} Delorme, P. et al. 2010, \aap, 518, A39
\bibitem[Eisenhardt et al.(2010)]{eis10} Eisenhardt, P. R. M. et al. 2010, \aj, 139, 2455
\bibitem[Faherty et al.(2009)]{fah09} Faherty, J., Burgasser, A., Cruz, K., Shara, M., Walter, F., \& Gelino, C.  2009, \aj, 137, 1
\bibitem[Kirkpatrick et al.(2005)]{kir05} Kirkpatrick J. D. et al., 2005, \araa, 43, 195
\bibitem[Kirkpatrick et al.(2010)]{kir10} Kirkpatrick J. D. et al., 2010, \apjs, 190, 100
\bibitem[Knapp et al.(2004)]{kna04} Knapp, G. R. et al. 2004, \apj, 127, 3553
\bibitem[Leggett et al.(2010)]{leg10} Leggett, S. K., Saumon, D., Cushing, M., Marley, M. \& Pinfield, D.  2010, \apj, 720, 252
\bibitem[L\'{e}pine et al.(2007)]{lep07} L\'{e}pine, S., Rich, R. M., \& Shara, M. M.  2007, \apj, 669, 1235
\bibitem[Liu et al.(2011)]{liu11} Liu, M. C. et al. 2011, ArXiv:1103.0014
\bibitem[Low \& Lynden-Bell(1976)]{low76} Low, C. \& Lynden-Bell, D., 1976, \mnras, 176, 367
\bibitem[Lucas et al.(2010)]{luc10} Lucas, P. W. et al. 2010, \mnras, 408, L56
\bibitem[Luhman et al.(2011)]{luh11} Luhman, K. L., Burgasser, A. J., \& Bochanski J. J. 2011, ApJL, 730, 9
\bibitem[Mainzer et al.(2011)]{mai11} Mainzer, A. et al. 2011, \apj, 726, 30
\bibitem[McLean et al.(1993)]{mcl93} McLean, I. S. et al. 1993, Proc. SPIE, 1946, 513
\bibitem[Nakajima et al.(1995)]{nak95} Nakajima, T., Oppenheimer, B. R., Kulkarni, S. R., Golimowski, D. A., Matthews, K., \& Durrance, S. T.  1995, \nat, 378, 463
\bibitem[Reid et al.(1995)]{rei95} Reid, I., Hawley, S. L., \& Gizis, J. E. 1995, \apj, 110, 1838
\bibitem[Rodriguez et al.(2011)]{rod11} Rodriguez, D. R., Bessell, M. S., Zuckerman, B., \& Kastner, J. H. 2011, \apj, 727, 62
\bibitem[Rodriguez et al.(2011b)]{rod11b} Rodriguez, D. R., Zuckerman, B., Melis, C., \& Song, I. 2011, \apj, 732, L29
\bibitem[Shkolnik et al.(2011)]{shk11} Shkolnik, E. L., Liu, M. C., Reid, I. N., Dupuy, T., Weinberger, A. J., 2011, \apj, 727, 6
\bibitem[Song et al.(2003)]{song03} Song, I., Zuckerman, B., \& Bessell, M. S. 2003, \apj, 599, 342
\bibitem[Vrba et al.(2004)]{vrba04} Vrba F. J. et al., 2004, \apj, 127, 2948
\bibitem[Warren et al.(2007)]{war07} Warren, S. J. et al. 2007, \mnras, 381, 1400
\bibitem[Wright et al.(2010)]{wri10} Wright E. L. et al., 2010, \aj, 140, 1868
\bibitem[van Leeuwen(2007)]{van07} van Leeuwen, F. 2007, in Astrophys. Space Sci. Libr. 350, $Hipparcos$: the New Reduction of the Raw Data
\bibitem[Zuckerman \& Song(2004)]{zuc04} Zuckerman, B. \& Song, I., 2004, \araa, 42, 685
\bibitem[Zuckerman \& Song(2009)]{zuc09} Zuckerman, B. \& Song, I., 2009, \aap, 493, 1149
\end{thebibliography}
\end{document}